\documentclass[showpacs,amsfonts,amsmath,nofootinbib,twocolumn,prl]{revtex4}%

 \usepackage{graphicx}
 \usepackage{dcolumn}
 \usepackage{bm}

\begin{document}
 \title{Measures and Mismeasures of Scientific Quality}
  \date{22 December 2005}
  \author{Sune Lehmann}
   \email{lehmann@nbi.dk}
   \affiliation{Informatics and Mathematical Modeling, Technical University of
   Denmark, Building 321, DK-2800 Kgs.~Lyngby, Denmark.}
  \author{Andrew D.~Jackson}
   \affiliation{The Niels Bohr Institute, Blegdamsvej 17, DK-2100 K\o benhavn \O, Denmark.}
  \author{Benny Lautrup}
   \affiliation{The Niels Bohr Institute, Blegdamsvej 17, DK-2100 K\o benhavn \O, Denmark.}
\begin{abstract}
We present a general Bayesian method for quantifying the statistical
reliability of one-dimensional measures of scientific quality based
on citation data. Two quality measures used in practice --- ``papers
per year'' and ``Hirsch's $h$'' --- are shown to lack the accuracy
and precision necessary to be useful. The mean, median and maximum
number of citations are reliable and permit accurate predictions of
future author performance on the basis of as few as 50 publications.
\end{abstract}

\maketitle

Although quantifying the quality of individual scientists is a
difficult task, most scientists would agree that: (i) it is better
to publish a large number of articles than a small number, and (ii)
for any given paper, its citation count (relative to citation habits
in its field) provides a useful measure of its quality.  Even given
the assumption that the quality of a scientist is related to his/her
citation record, it is still necessary to convert the details of a
citation record into an intensive (i.e., time-independent) scalar
measure of quality.  The questions of which measure of quality is
best and whether any such measure can be useful remain unanswered.
Nevertheless, a variety of measures of quality based on citation
data have been proposed in the literature and some have been adopted
in practice \cite{garfield:77,hirsch:05,repp:05}.  Their merits rely
largely on intuitive arguments and value judgments. The absence of
quantitative support for measures of quality based on citation data
is a matter of genuine concern since citation data is routinely
considered in matters of appointment and promotion which affect
every working scientist.

The purpose of analyzing and comparing citation records is to
discriminate between scientists. Any ranking is based on a single
real number $m$, presumed to be a quantitative measure of the
quality of a scientist's production.  Whatever the intrinsic and
value-based merits of this measure, it will be of no practical value
unless the corresponding uncertainty in its assignment is small.
From this point of view, the ``best'' choice of measure will be that
which provides maximal discrimination between scientists and hence
certainty in the values assigned.  The present paper is intended to
demonstrate that the question of deciding which of several proposed
measures is most discriminating, and therefore ``best'', can be
addressed quantitatively using standard Bayesian statistical
methods.

The present analysis is based on data from the
SPIRES\footnote{SPIRES contains virtually all papers in high energy
physics written since 1974 and their lists of references
\cite{spires}.} database of papers in high energy physics. Our data
set consists of all citable papers from the theory subfield, ultimo
2003, with all citations to papers outside of SPIRES removed. In
\cite{lehmann:03}, we have shown that the theory subsection of
SPIRES is a homogeneous data set\footnote{Citation distributions in
the ``Review'' and ``Instrumentation'' subsets are markedly
different.}. For the same reason we include only the publications of
``academic scientists'', defined as those with $25$ or more
published papers, and exclude those who cease active journal
publication early in their careers (see~\cite{lehmann:03a},
chapters~3 and~4).  The resulting data set includes $5\,787$ authors
and $282\,204$ papers. The actual number of papers is smaller since
multiple author papers are counted once per co-author, in agreement
with normal practice in publicly available citation counts
\cite{spires}. Note that the number of co-authors is relatively
small in this subfield (typically $1$--$3$ per theoretical paper),
and the effects of weighting papers by the number of co-authors have
been shown to be negligible \cite{supplemental}.

Like other sets of citation data, the data in this subset of SPIRES
is well-described by an asymptotic power law.  Specifically, the
probability that a paper will receive $n$ citations is approximately
proportional to $(n+1)^{-\gamma}$ with $\gamma = 1.10$ for $n \le
50$ and $\gamma = 2.70$ for $n > 50$.  The transition between these
two power laws is found to be quite sharp~\cite{lehmann:03}. As a
result, there is a significant difference between the mean of
$\approx$$18.4$ and median of $\approx$$5$ citations per paper. Note
that all higher moments of this distribution are ill-defined. This
alerts us to the possibility that the results of citation analyses
can depend sensitively on the chosen scalar measure of author
quality. The rationale underlying all citation analyses is that
citation data is strongly correlated such that a ``good'' scientist
has a far higher probability of writing a good (i.e., highly cited)
paper than a ``poor'' scientist. This expectation is fulfilled in
practice, and the citation data from SPIRES contain significant
longitudinal correlations~\cite{lehmann:03,lehmann:05}.

We thus categorize authors by a tentative quality index, $m$,
derived from their citation record. Once assigned, we can construct
the prior distribution, $p(m)$, that an author has measure $m$ and
the conditional probabilities, $P ( n | m )$, that a paper written
by an author with measure $m$ will receive $n$ citations. Studies
performed on the first $25$, first $50$ and all papers of authors
with a given value of $m$~\cite{lehmann:03} indicate the absence of
temporal correlations in the citation distributions of individual
authors.  In practice, we bin authors in deciles according to their
value of $m$ and papers logarithmically, due to the asymptotic power
law behavior noted above.  We have confirmed that the results here
are insensitive to binning effects.

We will consider six possible intensive measures of author quality.
Five of these  have been proposed and used in the literature. They
include the mean and median number of citations per paper, the
number of citations of an author's maximally cited paper, the number
of papers published per year, and a measure recently proposed by
Hirsch.\footnote{ Hirsch's definition is as follows: ``A scientist
has index $h$ if $h$ of his/her $N_p$ papers have at least $h$
citations each and the remaining $(N_p-h)$ papers have fewer than
$h$ citations each.''\cite{hirsch:05}. To obtain an intensive
measure, we adopt Hirsch's assumption that $h$ grows approximately
linearly with time and normalize each $h$ by the author's
professional age, defined as the time between the publication of
first and last papers.}  As a control of the statistical methods
adopted, we also consider the results of binning authors
alphabetically since an author's citation record should provide us
with no information regarding the author's name.

Each of these measures has disadvantages.  Since the average number
of citations is based on a finite sample drawn from a power-law
distribution, the addition or removal of a single highly cited paper
can materially alter an author's mean, cf. \cite{newman:05a}.
Although it is thus potentially statistically unreliable, the mean
is the most commonly used measure of author quality. This
reservation applies with even greater force if $m$ is the number of
citations of an author's single most highly cited paper.  In
addition, this measure cannot decrease with time and is not
guaranteed to be intensive for a currently active scientist.
Nevertheless, it is perfectly tenable to claim that the author of a
single paper with $1000$ citations is of greater value to science
than the author of $10$ papers with $100$ citations each even though
the latter is far less probable for power-law distributions. The
maximally cited paper might provide better discrimination between
authors of ``high'' and ``highest'' quality, and this measure merits
consideration. Alternatively, one can measure excellence by the
median number of citations of an author's papers. In contrast to
mean and maximum citations, the median is statistically robust. The
median (or any other percentile) of $\mathcal{N}$ random draws on
\emph{any} normalized probability distribution is Gaussian
distributed in the limit $\mathcal{N}\to\infty$ \cite{supplemental}.
While the statistical stability of the median (and percentiles)
makes it well-suited for dealing with power laws, reservations can
again be expressed. The democratic use of all data points tends to
ignore the possibility that an author's true merit lies in the most
highly cited papers. Another widely used measure of scientific
quality is the average number of papers published by an author per
year.  This would be a good measure if all papers were cited equally
or if all papers were of equal scientific merit.  The data make it
clear that scientific papers are not cited equally, and few
scientists hold the view that all published papers are of equal
quality and importance. Roughly 50\% of all papers in SPIRES are in
fact cited less than 2 times including self-citation. Indeed, if all
papers were of equal merit, citation analyses would provide a
measure of industry rather than intrinsic quality!

Finally, Hirsch's measure attempts to find a balance between
productivity and quality and to avoid the heavy weight which
power-law distributions place on a relatively small number of highly
cited papers. As with other such attempts (e.g., the median), it can
lead to anomalous measures at the high end of the scale. More
seriously, Hirsch establishes an equality between incommensurable
quantities. An author's papers are listed in order of decreasing
citations with paper $i$ have $C(i)$ citations. Hirsch's measure is
determined by the equality, $h = C(h)$, of two quantities with no
evident logical connection.  While it might be reasonable to assume
that $h \sim C(h)^{\kappa}$, there is no reason why both $\kappa$
and the constant of proportionality should be precisely $1$.

We have binned the SPIRES authors and their citation records
according to each of the six tentative measures, $m$, above
\cite{supplemental}. We have constructed the prior distribution, $p
( \alpha )$, that an author is in author bin $\alpha$ and the
conditional probability, $P( i | \alpha )$ that a paper by an author
in bin $\alpha$ will fall in citation bin $i$. We now wish to
calculate the probability, $P ( \{ n_i \} | \alpha )$, that an
author in bin $\alpha$ will have a citation record with $n_i$ papers
in each citation bin. To do this, we assume that citations for the
$M$ papers written by a given author with $n_i$ papers in citation
bin $i$ are obtained from $M$ independent random draws on the
appropriate distribution, $P( i | \alpha  )$.  Thus,
\begin{equation}\label{eq:conditionalindependence}
    P(\{n_i \}| \alpha ) = M! \, \prod_{i} \,  \frac{P( i| \alpha )^{n_i}}{(n_i)!} \ .
\end{equation}
We have already noted the absence of large-scale temporal variations
in $P ( i | \alpha )$ during an author's scientific life.  Other
correlations could be present.  For example, one particularly
well-cited paper could lead to an increased probability of high
citations for its immediate successor(s).  While it is difficult to
demonstrate the presence or absence of such correlations, the
results below provide a posteriori indications that such
correlations, if present, are not overly important.  We can invert
the probability $P (\{n_i\}| \alpha )$ using Bayes' Theorem to
obtain
\begin{eqnarray}
P(\alpha |\{n_i\})  &=& \frac{P(\{n_i\}| \alpha ) \, p( \alpha )}{p(\{n_i\})}\nonumber\\%
                     &=& \frac{p( \alpha ) \prod_{k} \, P( k| \alpha)^{n_k}}{\sum_{\alpha'}
                      p( \alpha' ) \, \prod_{k'} \, P(k'| \alpha' )^{n_{k'}}} \ .
\label{eq:bayes}%
\end{eqnarray}
Note that the combinatoric factors cancel.

The quantity $P ( \alpha | \{ n_i \} )$, which represents the
probability that an author with citation record $\{ n_i \}$ belongs
in quality bin (i.e., decile) $\alpha$, is of primary interest.
While any given measure (e.g., the mean number of citations per
paper) can be calculated immediately from an author's publication record,
the calculated values of $P ( \alpha  | \{ n_i \} )$ provide more
detailed and reliable information.  By exploiting differences
between the various conditional probabilities, $P ( \{ n_i \} | \alpha
)$, as a function of $\alpha$, eq.~(\ref{eq:bayes}) determines the
appropriate decile value of $m$ (or its most probable value) using
all statistical information in the data. The large fluctuations
which can be encountered in identifying authors by their mean
citation rate or by their maximally cited paper are thereby
materially reduced.  Further, by providing us with values of $P (
\alpha  | \{ n_i \} )$ for all $\alpha$, we have a statistically
trustworthy gauge of whether the resulting uncertainties in
the assigned value of $m$ are sufficiently small for it to be
a reliable measure of author quality.

In short, eq.~(\ref{eq:bayes}) provides us with a measure of an
author's expected lifetime quality along with information which
allows us to assess the reliability of this determination.
Obviously, the confidence with which we can assign a value of $m$
increases dramatically with the total number of published papers.
As we shall see, it is also sensitive to the quality measure chosen.
Measures of quality are of value only to the extent that they can be
assigned to individual authors with high confidence.  The methods
described above allow us to determine this confidence for any choice
of measure in a manner which is value-free and completely
quantitative.

\begin{figure}[htbf]
 \begin{tabular}{ccc}
  \emph{(a)} First Initial &\emph{(b)} Papers/year &\emph{(c)} Hirsch\\
  \includegraphics[width=.3\hsize]{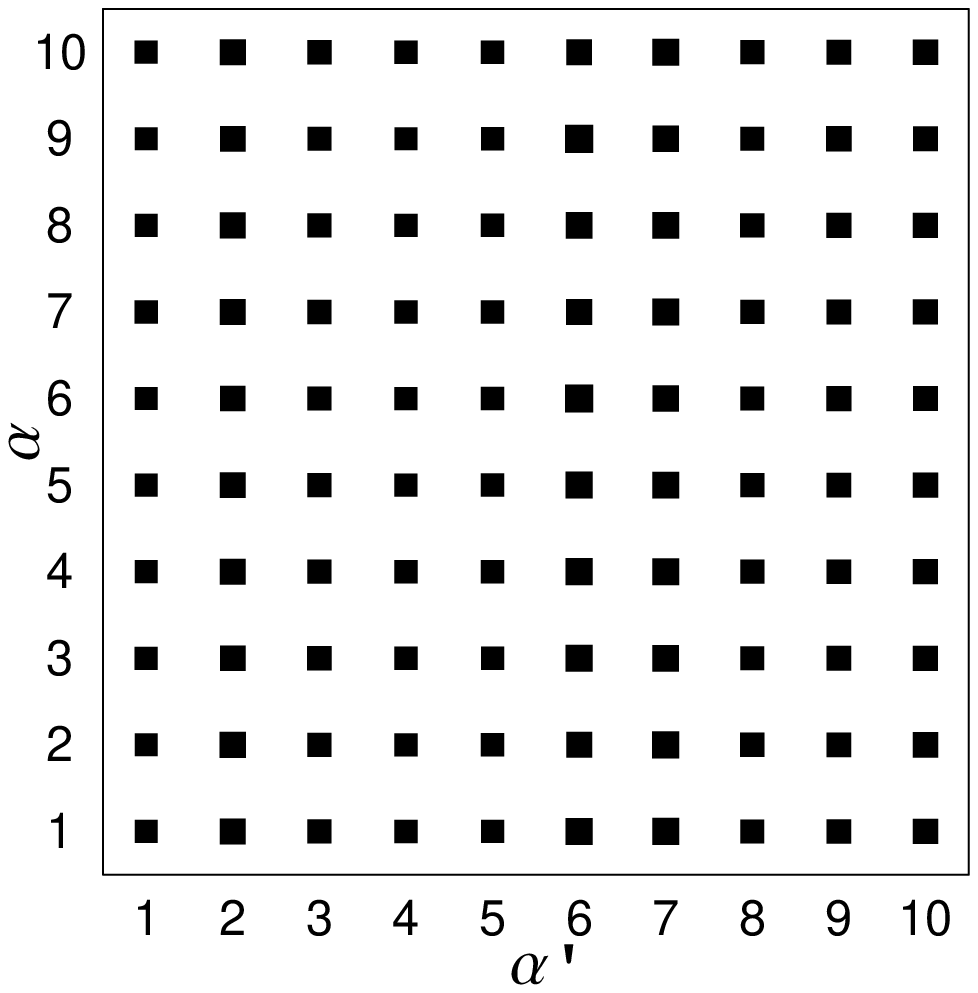} &%
  \includegraphics[width=.3\hsize]{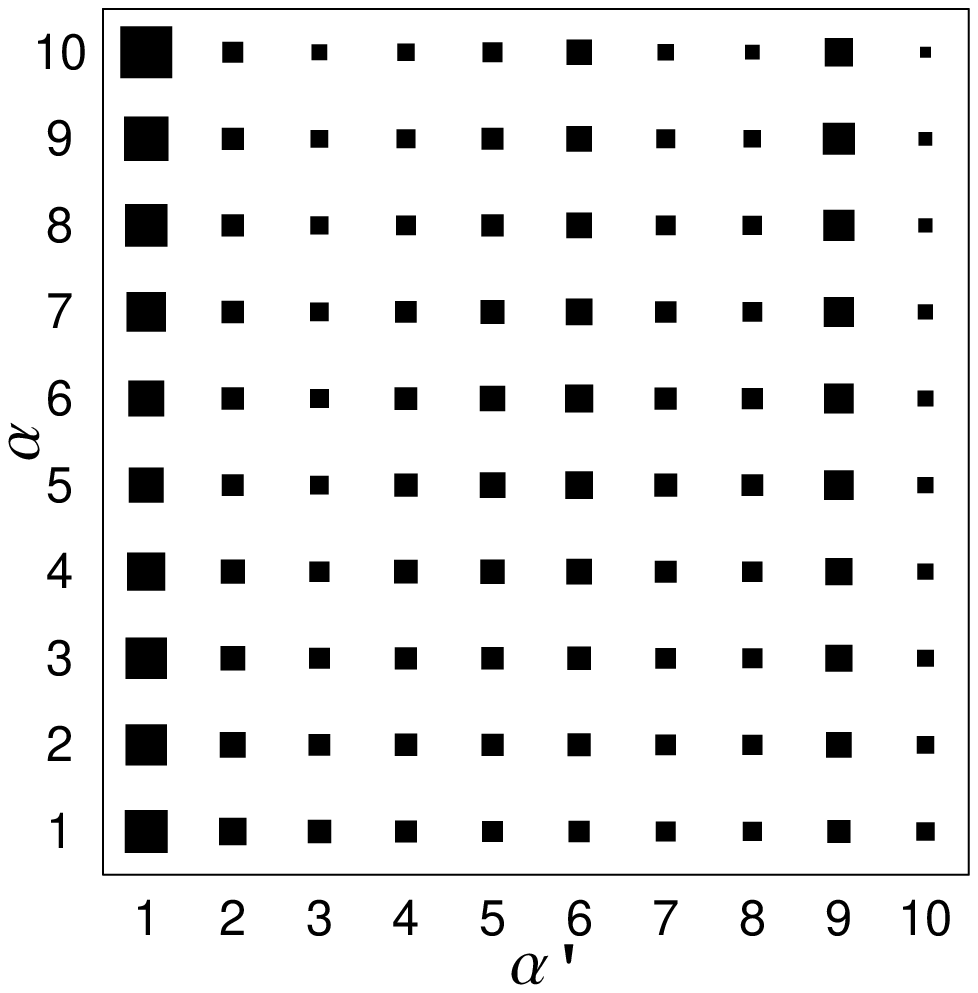} &%
  \includegraphics[width=.3\hsize]{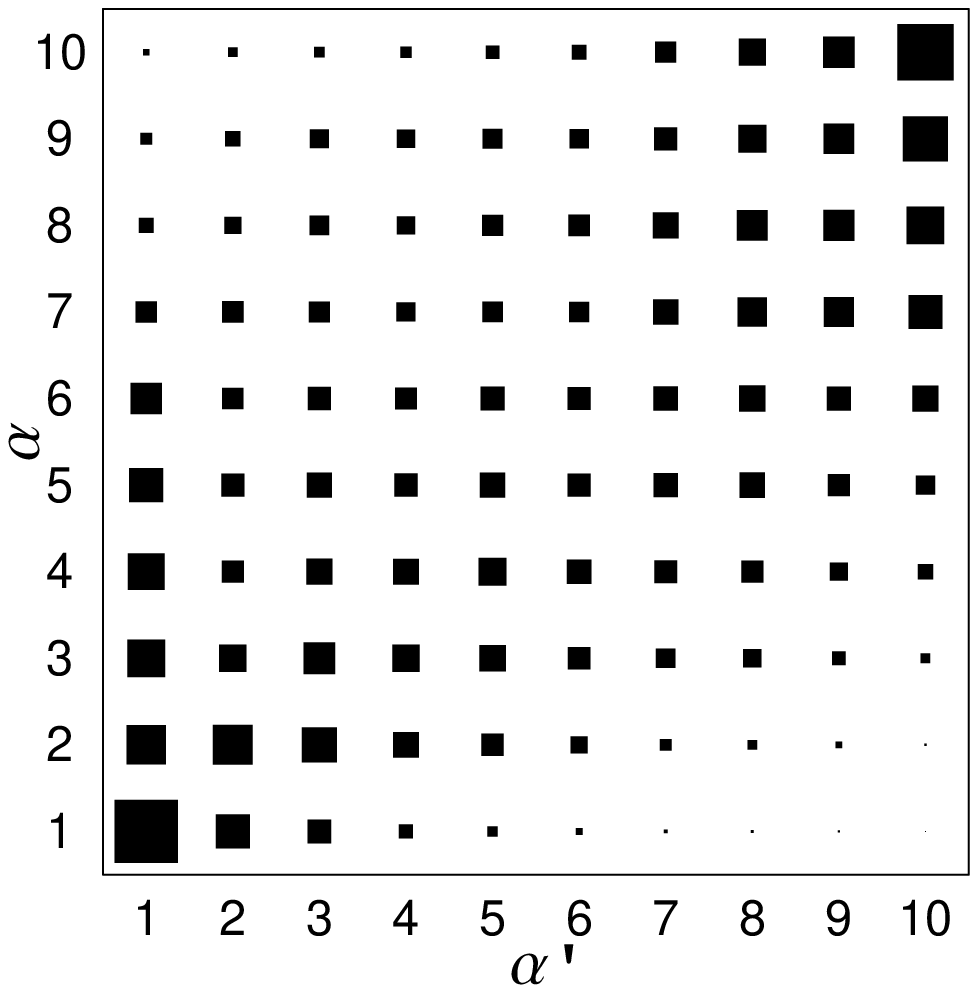}\\
  \emph{(d)} Mean &\emph{(e)} Median &\emph{(f)} Max\\
  \includegraphics[width=.3\hsize]{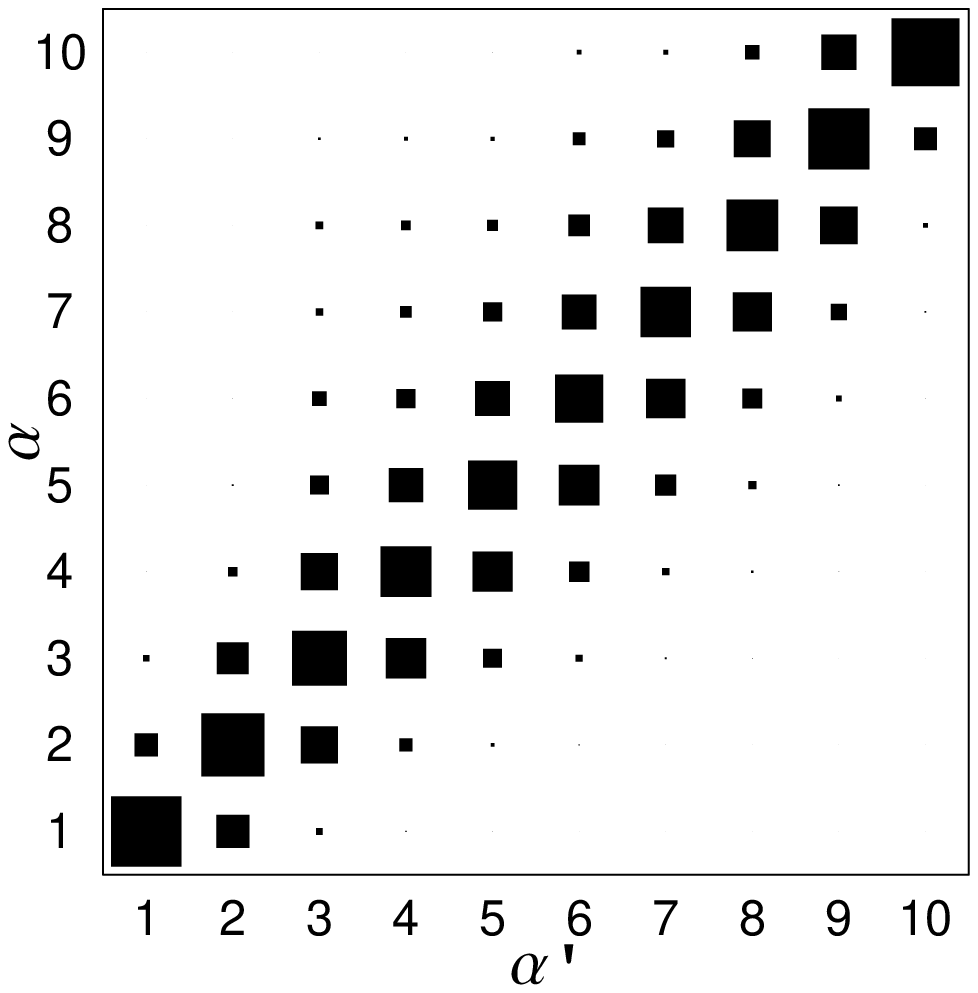} &%
  \includegraphics[width=.3\hsize]{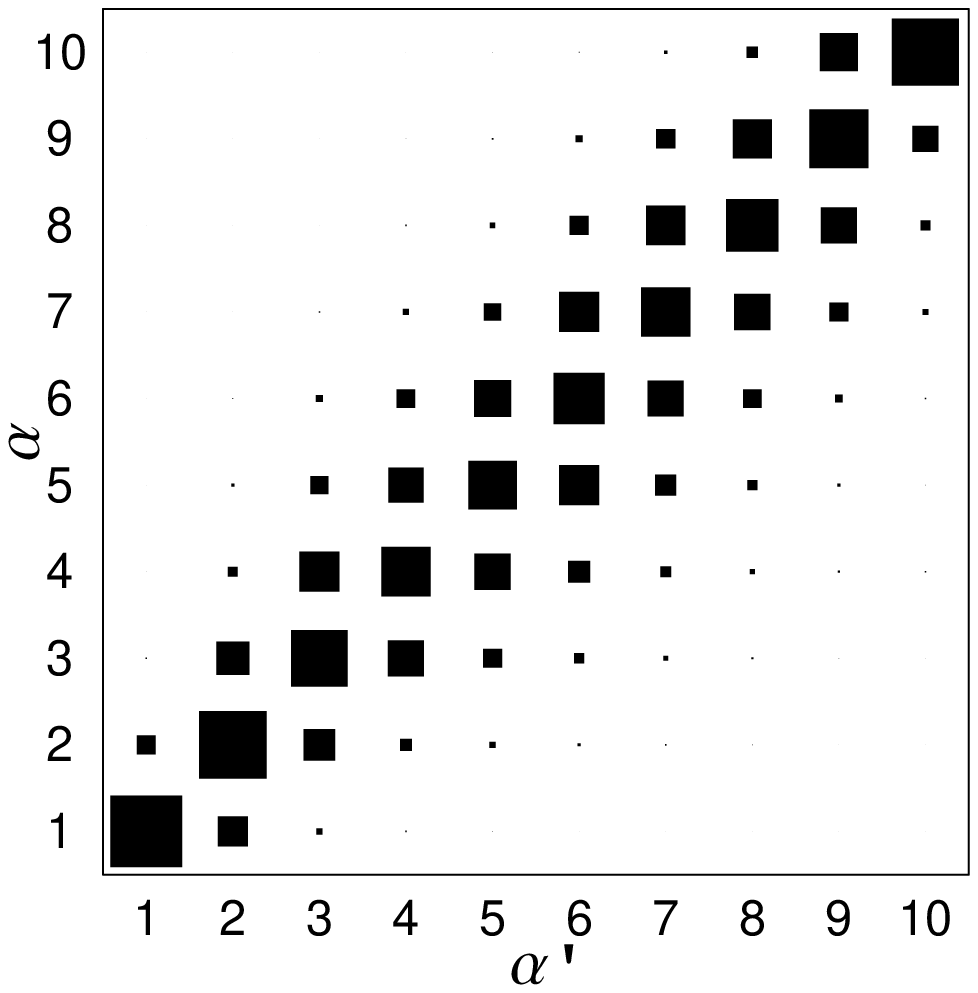} &%
  \includegraphics[width=.3\hsize]{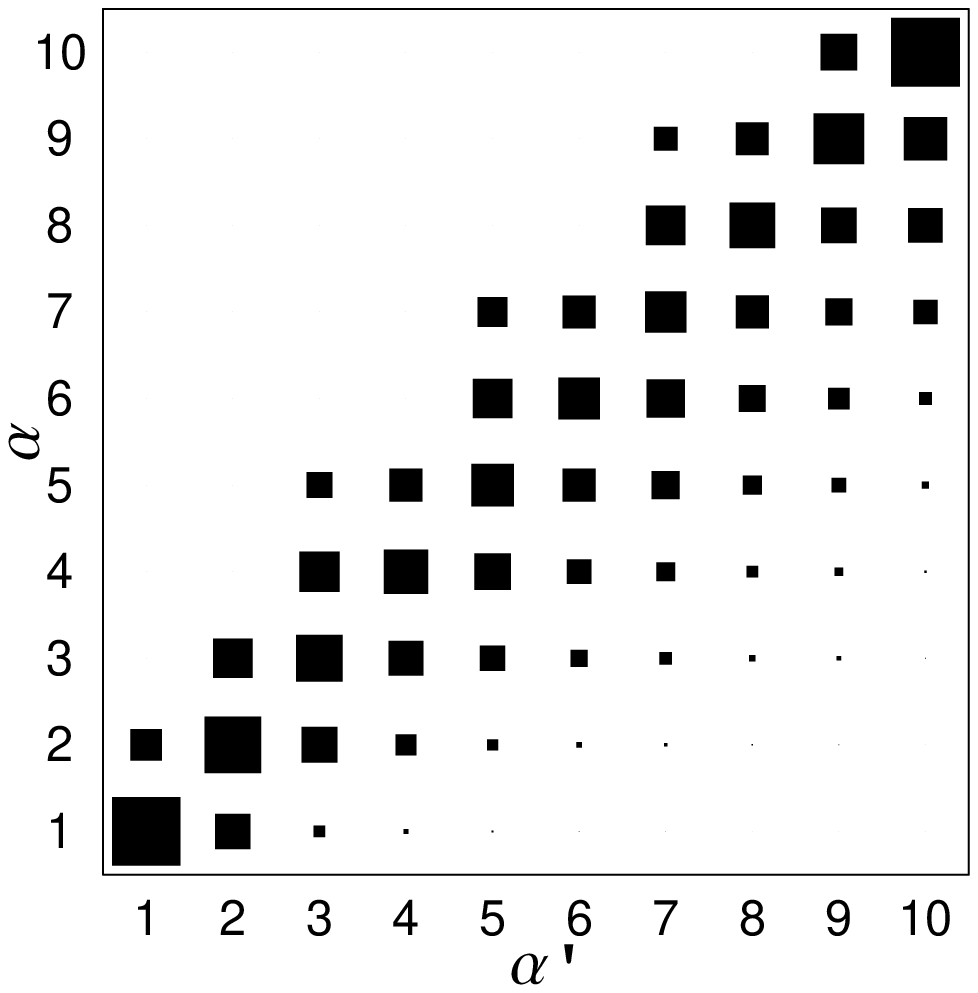}
 \end{tabular}
\caption{The probabilities, $P ( \alpha' | \alpha )$, for six
different measures. Each horizontal row, indexed by $\alpha$, shows
the average probabilities that authors initially assigned to a given
decile bin $\alpha$ are predicted to lie in the various decile bins
$\alpha'$. These probabilities are proportional the areas of the
corresponding squares.} \label{fig:avePnms}
\end{figure}

We now wish to explore the utility of each of the six measures
introduced above. To do this, we use Eq.~(\ref{eq:bayes}) to
calculate the probabilities, $P ( \alpha' | \{ n^{(\mu)}_i \} )$,
that each author, $\mu$, in SPIRES assigned to bin $\alpha$ by
direct measurement, is predicted to lie in bin $\alpha'$.   We then
construct the average probability, $P ( \alpha' | \alpha  )$, as the
simple average of the $P ( \alpha' | \{ n^{(\mu)}_i \} )$ over all
authors $\mu$ in bin $\alpha$. The results are shown ``stacked'' in
Fig.~\ref{fig:avePnms} for the various measures of excellence
considered.  Here, the $j$th horizontal row in each frame shows the
probabilities than an author initially assigned to decile $\alpha$
is predicted to be in decile $\alpha'$ by Eq.~(\ref{eq:bayes}).
This probability is proportional to the area of the corresponding
squares.  A perfect quality measure would place all weight in the
diagonal entries of these plots. Weights should be centered about
the diagonal for an accurate identification of author quality and
the certainty of this identification grows as more weight
accumulates in the diagonal boxes.  Note that the assignment of a
measure, e.g., the median citation rate, on the basis of
Eq.~(\ref{eq:bayes}) for any given author is likely to be more
accurate than the value obtained by direct computation since the
former is based on all information contained in the citation record.

All three measures shown in the bottom row of the figure perform
well. The maximum measure tends to overestimate an author's initial
decile assignment. This is understandable since the production of a
single paper with citations in excess of the values contained in bin
$\alpha'$ necessarily implies that the probability that he will lie
in this bin is $0$. The fact that the probabilities for these bins
shown in Fig.~\ref{fig:avePnms} are not strictly $0$ is a
consequence of the use of finite bin sizes.  The figure also makes
it clear the `first initial' measure fails both with regard to
accuracy and precision. The near constancy of $P( \alpha' | \alpha
)$ seen in this panel is expected for any random binning of authors
which ignores statistically natural groupings \cite{supplemental}.
The `publications per year' measure also fails both with regard to
accuracy and precision.  The dominant role played by individual
vertical columns and the fact that $P( \alpha' | \alpha )$ is
approximately independent of $\alpha$ is characteristic of schemes
which bin authors in a fashion that is systematic but inconsistent
with genuine correlations in the system.  In spite of a slight trend
towards the diagonal, similar criticism can be made of Hirsch's
measure (normalized as described above).  The median appears to be
the most balanced of the measures considered.

There are a variety of ways to assign numerical uncertainties to the
results shown in the figure.  For the good measures in the bottom
row, it is sensible to consider the average percentile assignment and
its rms uncertainty.  Using the median, we thus conclude that
authors in the ninth bin lie in the $82 \pm 8$ percentile on average.
Since such estimates convey little information about the ``mismeasures''
shown in the top row, it can be better to consider the entropy of
these predictions defined as
\begin{equation}
S = - \sum_{\alpha , \alpha'} \, P(\alpha' | \alpha) {\rm log}_2 \left[
P(\alpha' | \alpha) \right] p ( \alpha ) \ .
\label{entropy}
\end{equation}
This entropy has a minimum value of $0$ when $\alpha'$ is given
uniquely as a function of $\alpha$ to a maximum value of
$S_{\rm max} = {\rm log}_2 (10 )$ when all $P( \alpha |
\alpha' ) = 1/10$.  So defined, the entropy tells us the average
number of bits required to determine $\alpha'$ for a given $\alpha$.
Good measures correspond to small values of $S/S_{\rm max}$.  The
values of $S/S_{\rm max}$ are $0.998$, $0.919$, $0.855$, $0.509$,
$0.489$, and $0.583$ for the measures (a)--(e), respectively.

It is clear from eq.~(\ref{eq:bayes}) that the ability of a given
measure to discriminate is greatest when the differences between the
conditional probability distributions, $P(i | \alpha )$, for
different author bins $\alpha$ are greatest.  These differences can
quantified by measuring the ``distance'' between two such
conditional distributions with the aid of the Kullback-Leibler (KL)
divergence (also know as the relative entropy).  The KL divergence
between two discrete probability distributions, $p$ and $p'$,
is defined as
\begin{equation}\label{eq:kldef}
    \mathrm{KL}[p,p']=\sum_{i}p_i \log_2 \left(\frac{p_i}{p_i'}\right) \ .
\end{equation}
Calculation of the KL divergence for the conditional distributions
$P( i | \alpha )$ and $P( i | \alpha' )$ for the various quality
measures considered confirms the conclusions drawn from
Fig.~\ref{fig:avePnms} and from the values of $S/S_{\rm max}$.
Publication rate and Hirsch's $h$ (as well as alphabetization) fail
as useful measures of author quality; mean, median and maximal
citation rates are all successful and virtually equivalent measures.

Finally, we address the question of how many published papers are
needed to make a reliable prediction of the lifetime quality
measure for a given author. Here, we consider only results
using the median citation rate as a measure. If this number is
sufficiently small, analyses along the lines presented here can
provide a practical tool of potential value for predicting long-term
scientific accomplishment. To this end, we consider how $P( \alpha
|\{n_i\})$ scales with the total number of published papers, $M$,
for the most probable in bin $\alpha$ with $n_i = M P( i| \alpha )$.
Using eq.~(\ref{eq:bayes}), we obtain the general result that the
probability of assigning an average author to the wrong bin vanishes
exponentially as $M \to \infty$.  Given enough papers and a reliable
measure, the correct author bin will ultimately dominate. To correctly
assign the most probable to outer deciles 1, 2, 3 and 8, 9, 10 at
the $90$\% confidence level requires respectively  $M=10$, $40$, $50$,
$50$, $50$, and $30$ papers.

All quality measures have difficulty in making correct assignments
to deciles $4$--$7$. This apparent difficulty is due to our decision
to group authors by deciles.  It can be understood by assuming that
the distribution of intrinsic author quality has a maximum at some
non-zero value.  Such an assumption seems reasonable if we imagine
that Nature provides a high-end cutoff and academic appointment
procedures filter out the least able.  For any such distribution,
the probability density will be highest for authors in the vicinity
of this maximum. The binning of authors by deciles or percentiles
then invites us to make distinctions where no material quality
difference exists.  The results of Fig.~\ref{fig:avePnms} or
calculations of the KL divergence remind us that we cannot do so. On
the other hand, the probability that an author can be correctly
assigned to one of these middle bins on the basis of $50$
publications is high.

As emphasized in the introduction, there are two distinct questions
which must be addressed in any attempt to use citation data as an
indicator of author quality. The first is whether the measure chosen
to characterize a given citation distribution or even the
citation distribution itself truly reflects the qualities that we would
like to probe.  The second is whether a given measure is capable of
discriminating between authors in a reliable fashion and, by
extension, which of several measures discriminates best.  We have
shown that the use of Bayesian statistics makes it possible to
answer this second question in a value-neutral and statistically
compelling manner.  We have thus shown that alphabetization,
papers per year, and Hirsch's measure fail to provide a faithful
scalar measure of full citation records and cannot
be regarded as useful measures of author quality.  The situation
is quite different for the mean, median and maximum citation measures.
They all lead to reliable conclusions regarding an author's citation
record on the basis of $\approx 50$ published papers, and it is possible
to assign meaningful statistical uncertainties to the results. Further, the generally
high level of discrimination found with these measures provides indirect
support for our assumption that there are no additional correlations of material
importance in the citation data, so that an author's citation record
can be regarded as obtained from a random draw on the appropriate
conditional distribution, $P( i | \alpha )$. The difficulty
encountered in discriminating between authors in the middle deciles
suggests that intrinsic author ability is peaked about some non-zero
value.

Given homogeneous subsets of data, the methods presented here also
permit the meaningful comparison of scientists working in different
fields with minimal value judgments. It seems fair, for example, to
declare equality between a condensed matter experimentalist and a
high energy theorist provided that they are in the same percentile
of their respective peer groups.  Similarly, it is possible to
combine probabilities in order to assign a quality level to authors
with publications in several disjoint subfields. All that is
required is knowledge of the conditional probabilities for the
distribution of citations in each homogeneous subgroup. The fact
that roughly $50$ publications are sufficient to draw meaningful
conclusions about author quality suggests that the present methods
can provide information useful in the academic appointment process.
In this regard, we note that there are strong indications that the
initial publications of a given author are drawn (at random) on the
same conditional distribution as his/her remaining
papers~\cite{lehmann:03a}.  It is clear, however, that it takes time
for a paper to accumulate its full complement of citations.  While
this has not been taken into account here, present methods readily
permit its inclusion.  Subjecting citation data to more serious
statistical analysis can suggest new and potentially interesting
applications. For example, one practical hiring strategy would be
commitment to the principle that no new appointment should knowingly
lower the average (or median) quality of the department in question.
Finally, we note that, when unable to measure that which they would
like to maximize (e.g., quality), scientists are inclined to
maximize what they know how to measure.  The confidence with which
it can be assigned may not be the only criterion for selecting a
measure of scientific quality.  However, it can and should be
considered. The methods proposed here offer simple and reliable
tools appropriate for addressing all of these issues.

\bibliographystyle{unsrt}
\bibliography{newbib}

\begin{thebibliography}{1}
\bibitem{supplemental}
Data and further details are avalable as supporting material.
\end{thebibliography}

\end{document}